\documentclass[a4paper, 11pt]{article}
\usepackage{jinstpub}

\usepackage{graphicx}
\usepackage{textgreek}

\usepackage[sorting=none]{biblatex}
\newif\ifdraft
\newif\ifextrafigs

\def\dvers{v0.5b}

\ifdraft
\usepackage{lineno}
\linenumbers
\usepackage{fancyhdr}
\usepackage[long, 24hr]{datetime}
\else
\extrafigsfalse
\fi

\title{Application of the VMM ASIC for SiPM-based calorimetry}

\author[a]{I.~Bearden}
\author[b,*]{V.~Buchakchiev}
\author[a]{A.~Buhl}
\author[a]{L.~Dufke}
\author[c]{T.~Isidori}
\author[a]{S.~Jia}
\author[b]{V.~Kozhuharov}
\author[d]{C.~Loizides}
\author[e,f]{H.~Muller}
\author[e,g]{D.~Pfeiffer}
\author[h]{M.~Rauch}
\author[d,i]{A.~Rusu}
\author[b]{R.~Simeonov}

\emailAdd{$^*$valentin@phys.uni-sofia.bg}

\affiliation[a]{University of Copenhagen, Copenhagen, Denmark}
\affiliation[b]{Faculty of Physics, University of Sofia, Sofia, Bulgaria}
\affiliation[c]{The University of Kansas, Lawrence, USA}
\affiliation[d]{ORNL, Oak Ridge, USA}
\affiliation[e]{European Organization for Nuclear Research (CERN), 1211 Geneva 23, Switzerland}
\affiliation[f]{Physikalisches Institut, University of Bonn, Nu{\ss}allee 12, 53115 Bonn, Germany}
\affiliation[g]{European Spallation Source ERIC (ESS), Box 176, SE-221 00 Lund, Sweden}
\affiliation[h]{University of Bergen, Bergen, Norway}
\affiliation[i]{SRS Technology, 30 Promenade des Artisans, 1217 Meyrin, Switzerland}

\abstract{
    Highly integrated multichannel readout electronics is crucial in contemporary particle physics experiments. 
    A novel silicon photomultiplier readout system based on the VMM3a ASIC was developed,
    for the first time exploiting this chip for calorimetric purposes.
    To extend the dynamic range
    the signal from each SiPM channel was processed by two electronics channels with different gain. 
    A fully operational prototype system with 256 SiPM readout channels
    allowed the collection of data from a prototype of the ALICE Forward 
    Hadron Calorimeter (FoCal-H). The design and the test beam results using high energy hadron beams are presented and discussed, confirming the 
    applicability of VMM3a-based solutions for energy measurements in a high rate environment.
}

\begin{document}
 
\maketitle

\ifdraft
\pagestyle{fancyplain}
\fancyhead{}
\fancyhead[L,L]{\color{red}DRAFT \dvers}
\fancyhead[R,R]{\color{red}INTERNAL ONLY}
\fancyfoot[L,L]{\color{red}Compiled: \today}
\fancyfoot[R,R]{\color{red}at \currenttime\ UCT}
\fi

\section{Introduction}
\label{sec:Intro}
    ALICE FoCal~\cite{bib:FoCal_LOI, bib:FoCal_TDR} is a forward calorimeter which will be installed
    during the Long Shutdown 3 (LS3) of the Large Hadron Collider (LHC) 
    at CERN, and will start taking data during LHC Run~4. Several testing 
    campaigns addressed the design and the physics performance of 
    prototypes of the electromagnetic (FoCal-E) and the hadronic (FoCal-
    H) sections of FoCal~\cite{bib:FoCal_Performance}. In addition, various readout systems were used 
    with the FoCal-H prototypes. The present study addresses the 
    compatibility between the FoCal-H Prototype 2 and the SRS (Scalable 
    Readout System) with the VMM Hybrid front-end electronics.

    The SRS~\cite{bib:SRS, bib:SRS_Dev} is a versatile and highly 
    adaptable solution for data transfer between detectors and computers. 
    The system was developed by the RD51 Collaboration and allows use 
    with detectors of any size while also lowering costs due to only 
    part of the system requiring redevelopment for new applications.
    As part of the NSW (New 
    Small Wheel) upgrade on the ATLAS experiment, the SRS system was used 
    to implement  new specially developed ASIC, the VMM3a 
    ~\cite{bib:VMM_SRS, bib:VMM_SRS_Rate}, to replace the original APV25 
    readout chips~\cite{bib:NSW}.
    
    A VMM ASIC has 64 channels each with its own preamplifier, shaper, 
    peak detector and ADCs. The chip can be programmed for multiple 
    applications as its gain, polarity, peaking time, threshold and 
    timing precision are all adjustable. It was originally developed for 
    use with Micromega and GEM gaseous detectors and is tailored towards 
    high event rates with low channel activation rates per event. 
    FoCal-H prototype is designed as a plastic-absorber calorimeter 
    with a SiPM-based light detection. The expected event rate during 
    test beam campaigns is of the order of 10 kHz, but a large number of channels are 
    activated for each event. Due to the hadronic shower's geometry, most 
    of the activated channels are usually connected to a single readout 
    chip, leading to a high and non-uniform load. In this paper we present 
    the results of an R\&D study of the usage of the VMM with one of the 
    FoCal-H prototypes which may extend the applicability of the VMM 
    ASIC to SiPM-based calorimeters in general.

\section{Experimental setup}
\label{sec:Setup}
    The FoCal-H 
    prototype consists of 9 modules with dimensions 
    ${\rm 6.5\times6.5\times110}$~cm$^{\rm 3}$
    in a ${\rm 3\times3}$ arrangement. 
    Each module consists of ${\rm 24\times28}$
    copper tubes with scintillating fibers inside them, 
    with 4 tubes removed for the installation of supporting rods.
    Light detection is provided by Hamamatsu S13360-6025PE SiPMs.
    The 668 fibers of each module are split into bundles. The outer 8 modules of the prototype are split into 25 channels with ${\rm \sim}$27 fibers per channel. The central one which detects the most energy is split into 49 channels with ${\rm \sim}$14 fibers per channel. 

    \begin{figure}[htp!]
    \begin{center}
        \includegraphics[width = 0.85\textwidth]{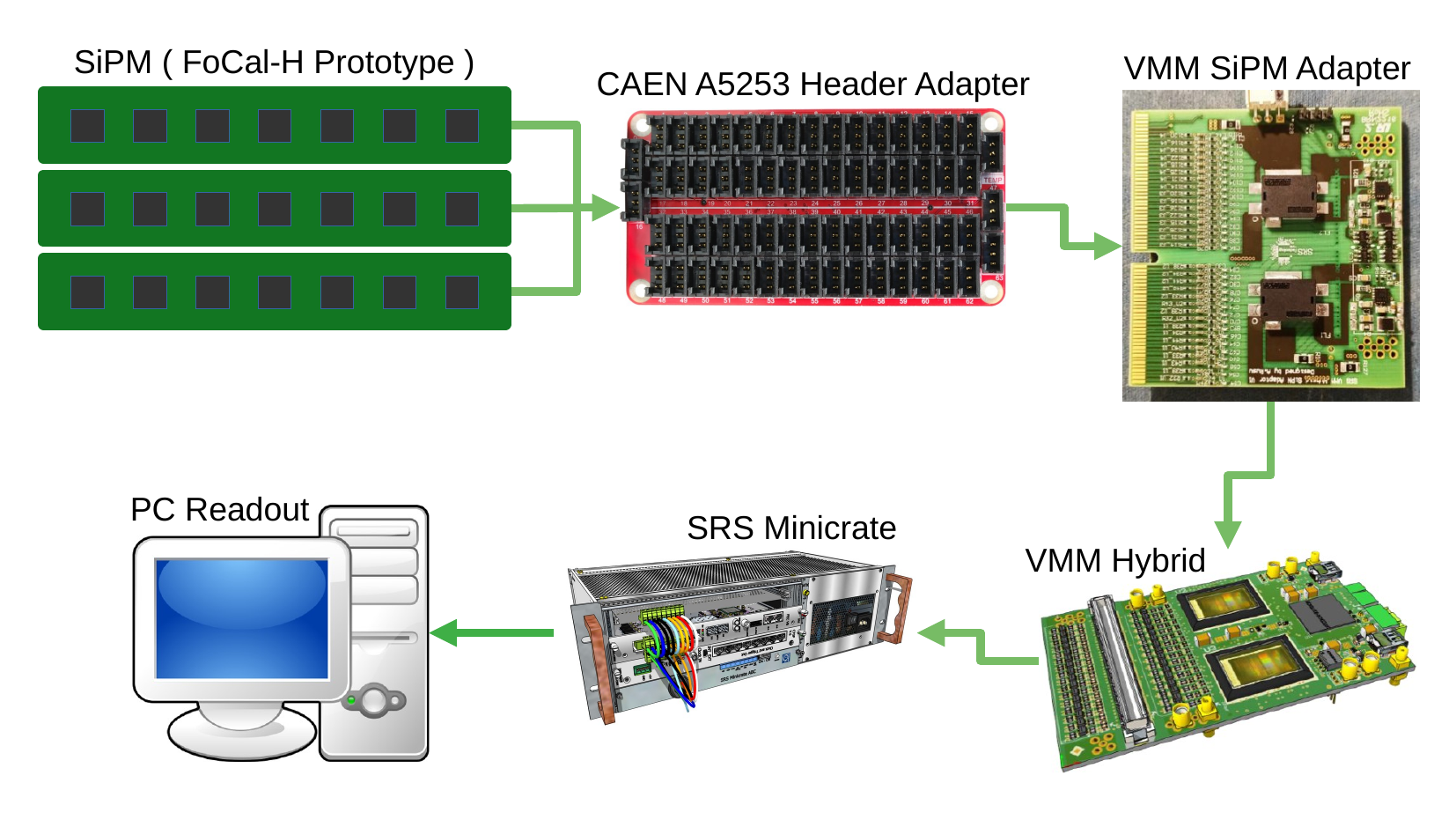}
        \caption{Schematic representation of the full readout setup.}
        \label{fig:Exp_setup}
    \end{center}
    \end{figure}

\begin{figure}[ht!]
    \begin{center}
        \begin{minipage}{0.49\textwidth}
            \includegraphics[width = \textwidth]{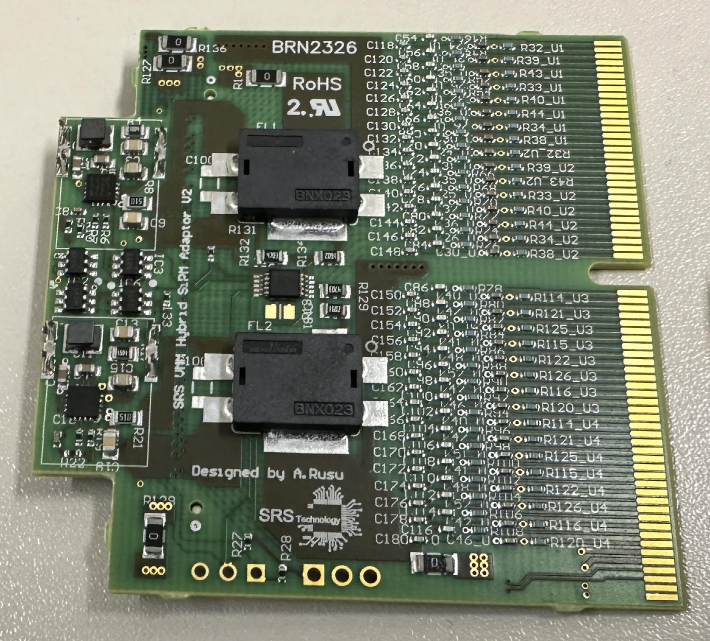}
        \end{minipage}
        \hspace{0.005\textwidth}
        \begin{minipage}{0.49\textwidth}
            \includegraphics[width = \textwidth]{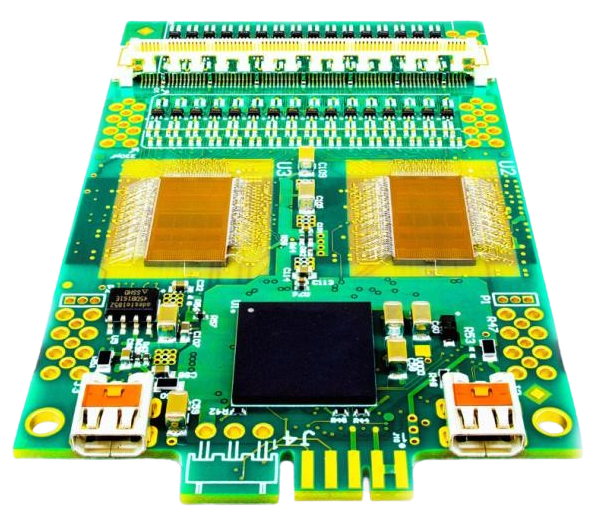}
        \end{minipage}
        \caption{Left - A photograph of the SiPM adapter board which allows the connection between the VMM Hybrid and the SiPMs through the CAEN A5253 header adapter. Right - The VMM Hybrid board which houses and configures the VMM ASICs.}
        \label{fig:SiPM_Adapter}
    \end{center}
    \end{figure}

    The FoCal-H prototype equipped with 4 VMM Hybrids was tested at the 
    SPS accelerator complex at CERN. The readout setup is presented in 
    Figure~\ref{fig:Exp_setup}. It consists of the following major components:

\begin{itemize}
    \item{The Focal-H prototype:} It is represented by the Silicon Photomultipliers. 
    
    \item{CAEN A5253 header adapter:} For facilitation, and independently of the 
    readout system, the SiPMs outputs from the FoCal-H back panel 
    were connected to the CAEN A5253 64 channel header 
    adapter~\cite{bib:CAEN_Header}.

    \item {The VMM SiPM adapter board:} To test the VMM readout
    solution with Silicon Photomultipliers, 
    a dedicated SiPM adapter board
    was developed
    in collaboration with RD51~\cite{bib:RD51_Coll}
    which can be plugged into the CAEN adapter.
    It acts as a 128 channel interface
    from the FoCal-H detector to the RD51 VMM hybrid. A prototype 
    was used in the FoCal test beam.

    \item{RD51 VMM hybrid:} The VMM hybrid board, 
    shown in figure \ref{fig:SiPM_Adapter}, right,
    houses two VMM3a ASICs, for a total of 128 electronics channels. 
    A Xilinx Spartan-6 Field Programmable
    Gate Array (FPGA) ensures the ASIC configuration and
    the communication (including hit data transfer) with 
    the SRS system via micro HDMI connectors~\cite{bib:Perf_RD51_VMM3a}. 

    \item{SRS minicrate:} The SRS back-end,
    as described in~\cite{bib:SRS},
    is contained in the SRS minicrate.
    It acts as a data aggregator and
    is able to serve up to 8 hybrid boards
    connected via HDMI interface.
    The SRS system also adds coarse timing information
    for time stamp calculations,
    and pads then feeds the data
    to the readout PC.
 
    \item{PC Readout:} A standard PC is used.
    The PC allows the configuration of the SRS readout by setting various parameters
    such as thresholds, gains, etc.
    The data, which is sent by the SRS system via a UDP protocol,
    is recorded by filtering the incoming network stream, and 
    further storing it on disk for processing. 
\end{itemize}
    
    The SiPM Adapter board is shown in Figure~\ref{fig:SiPM_Adapter}. 
    It transforms the 
    SIPM signal risetime into charge and houses 2 bias voltage generators which
    can provide between 0~V and 85~V (programmable trough an I2C interface).
    The HV distribution is split into 2 regions
    of 32 SIPMs which in the prototype, have the same bias voltage. The adapter uses a 
    precise DAC capable of adjusting the SiPM bias voltage 
    in steps of 5~mV. It also has an ADC to measure the actually provided 
    voltages and currents for the two regions.

\begin{figure}[ht!]
    \begin{center}
        \includegraphics[width = 0.85\textwidth]{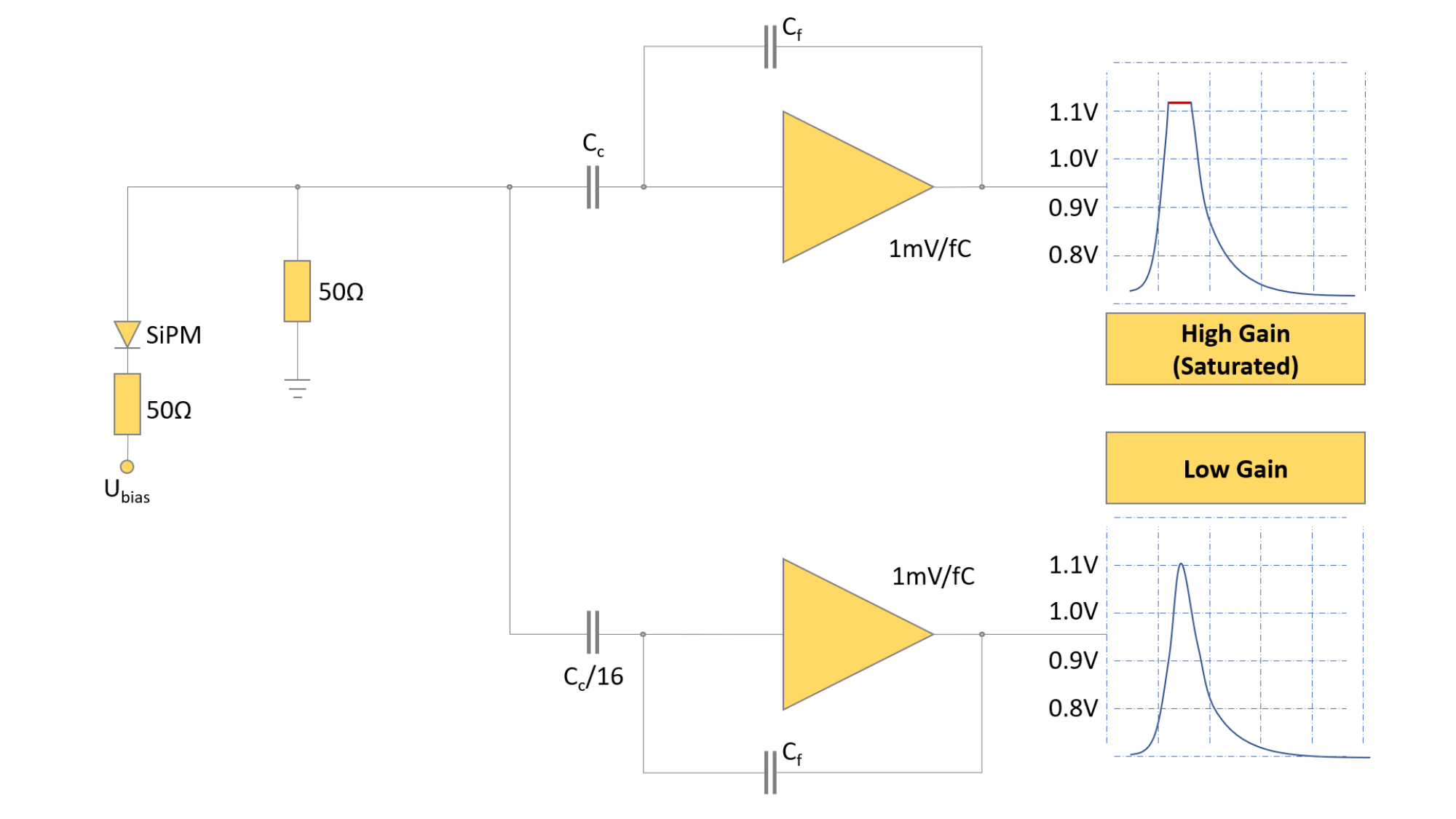}
        \caption{The circuit on the VMM SiPM adapter board which extends the dynamic range by splitting the SiPM signal into High Gain and Low Gain electronics channels.}
        \label{fig:HG/LG}
    \end{center}
    \end{figure}

    Initially, with the first prototype of the VMM SiPM adapter board, 
    only 32 channels per VMM (out of 64) were connected to the FoCal-H. 
    However, after testing the FoCal-H prototype with a hadron beam at 
    CERN SPS, due to  problems with ADC saturation, it was 
    decided to make use of the remaining 32 channels per VMM chip as 
    Low(er) Gain channels. 
    Therefore, the SiPM adapter board aims to extend the original dynamic 
    range of the VMM through the use of charge division coupling with 
    the same amplifier gain, resulting in  High Gain (HG) and Low 
    Gain (LG) channels. With this setup each SiPM channel is connected 
    to two VMM channels through identical amplifier circuits
    with the only exception being their coupling capacitors, as 
    can be seen in Figure~\ref{fig:HG/LG}.
    
    The chosen values of the capacitors are in a 16:1 ratio, 
    resulting in the same 16:1 ratio in the VMM output analog signals, 
    thus giving at least one non-saturated output for input signals with 
    much larger amplitudes.

\begin{figure}[ht!]
    \begin{center}
        \begin{minipage}{0.459\textwidth}
            \includegraphics[width = \textwidth]{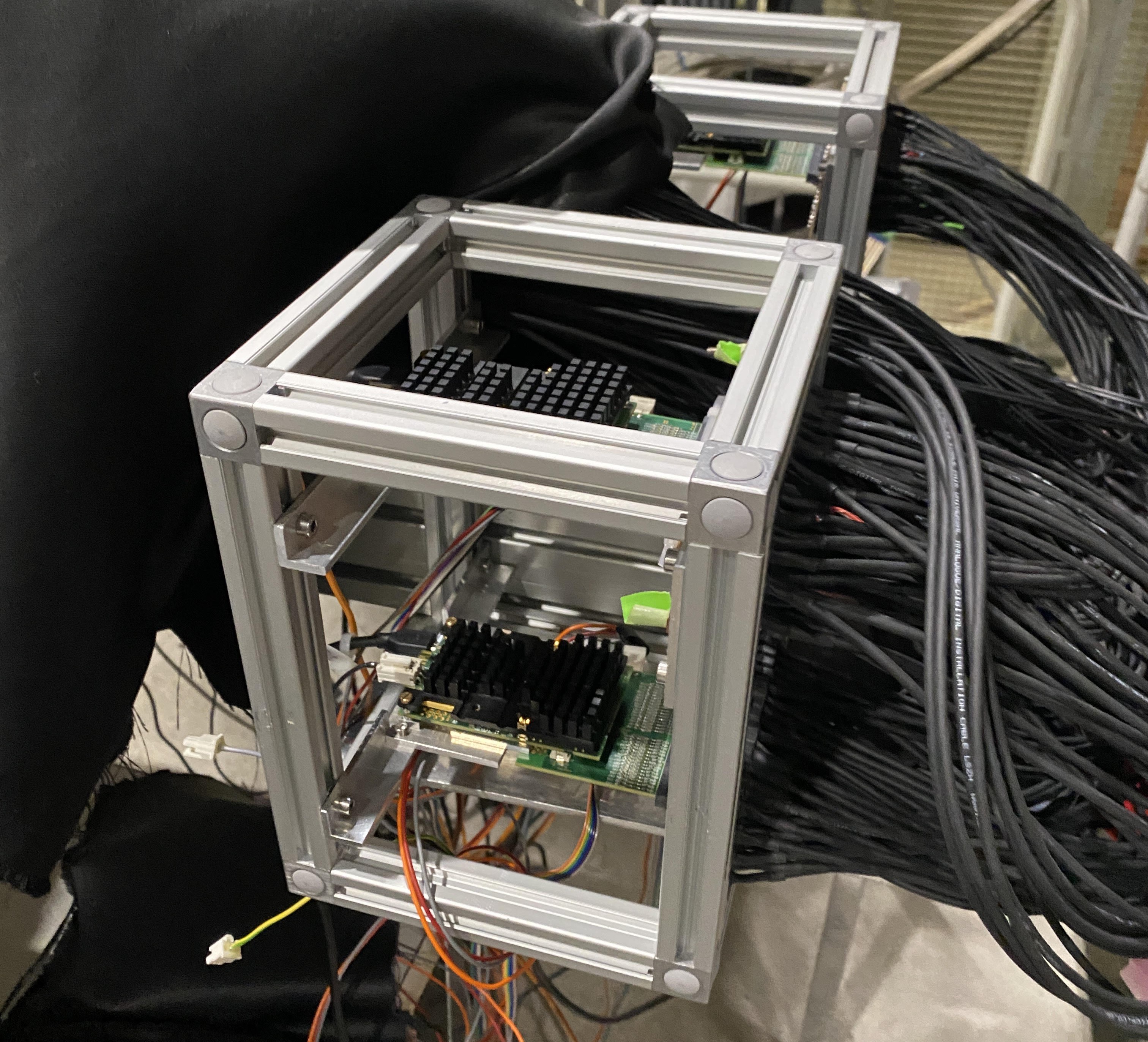}
        \end{minipage}
        \hspace{0.005\textwidth}
        \begin{minipage}{0.52\textwidth}
            \includegraphics[width = \textwidth]{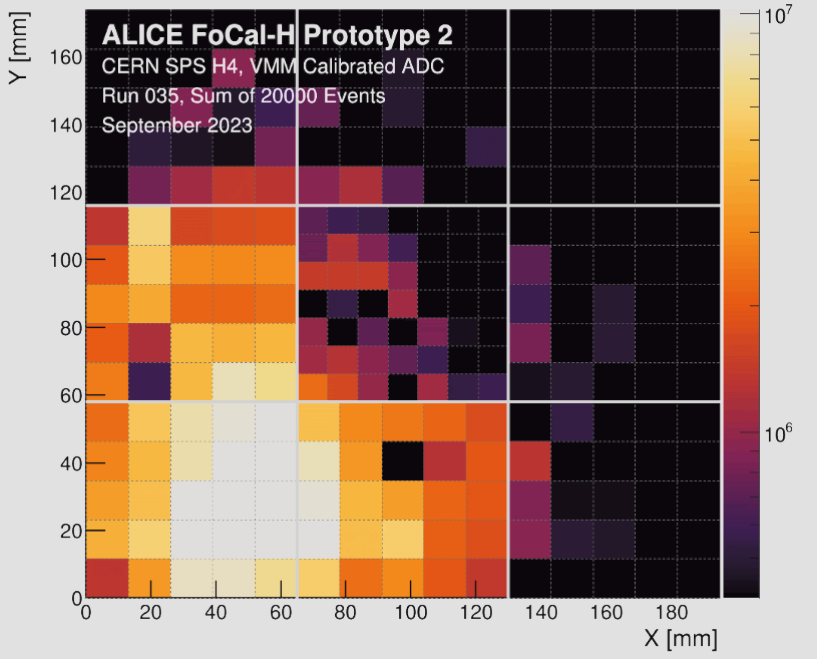}
        \end{minipage}
        \caption{ Left - VMM Hybrids connected to the back of the FoCal-H prototype through the SiPM adapter boards. Right - Cumulative charge in each FoCal-H channel in ADC units for 350~GeV beam energy.}
        \label{fig:VMM_connected}
    \end{center}
    \end{figure}

    Eight VMM chips were used to cover the 249 channels of the detector,
    with each detector channel being digitized by two VMM channels - one HG and one LG.
    The VMM connected to the back of the FoCal-H is presented in 
    Figure~\ref{fig:VMM_connected} along with a visualisation of the beam profile 
    using this setup with a 350~GeV hadron beam. 
    The detector was moved such that the center of the beam was inside the lower left 
    prototype module, as visible on the plot.

    Eighty-seven separate runs were performed using both hadron and 
    electron beams pointing towards the central module of the FoCal-H 
    prototype. The hadron beam energy was varied between 60~GeV and 350~GeV.
    Data for different gain and SiPM bias voltage was also collected
    for some of the available energies.

\section{Data analysis}
\label{sec:Analysis}
    A single data unit of the VMM is called a hit and is described by a 38 bit
    binary sequence in the stored data. The first 2 bits are flags. 
    The 3$^{\rm rd}$ through 8$^{\rm th}$ bits provide the channel ID (values 
    from 0 to 63); next 10 bits contain the digitized charge value; 8 
    bits a reserved for the TDC value and the last 12 bits - for the bunch-crossing ID (BCID)~\cite{bib:VMM_SRS_Rate}.
    
    The time stamp for each hit is calculated based on the BCID, the TDC and 
    the so-called FEC markers as described in~\cite{bib:VMM_SRS_Rate}.
    Afterwards, the stream is processed by bunching hits within an 8~
    $\mu$ s time window. Each such cluster is defined as a single event. 
    The total reconstructed charge is computed defined as the sum total ADC units 
    for all activated HG channels in the event:
        \begin{equation}
            Q^{\rm HG}_{\rm event} = \sum_i^N{{\rm ADC}_i^{\rm HG}}
        \end{equation}
    where $N$ is the number of the in-time hits. In order to suppress 
    noise, events with $N<6$ are rejected.

\begin{figure}[ht!]
    \begin{center}
        \hspace{1.25 mm}
        \includegraphics[width = 0.815\textwidth]{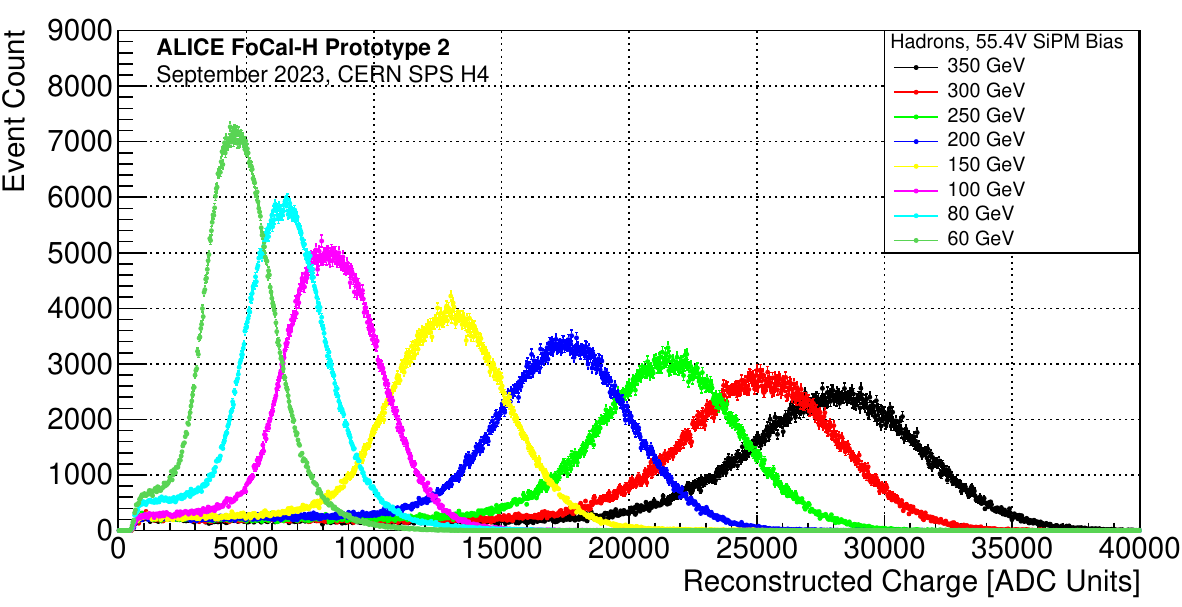}
        \caption{Comparison of the total charge distributions for all available beam energies between 60~GeV and 350~GeV. Distributions are normalised by the total number of events. SiPM bias voltage set to 55.4~V.}
        \label{fig:en_recon} 
    \end{center}
    \end{figure}

    The distributions of $Q^{\rm HG}_{\rm event}$ in ADC units for data 
    collected with different beam energies is presented in 
    Figure~\ref{fig:en_recon}.

    \begin{figure}[ht!]
    \begin{center}
        \includegraphics[width = 0.8\textwidth]{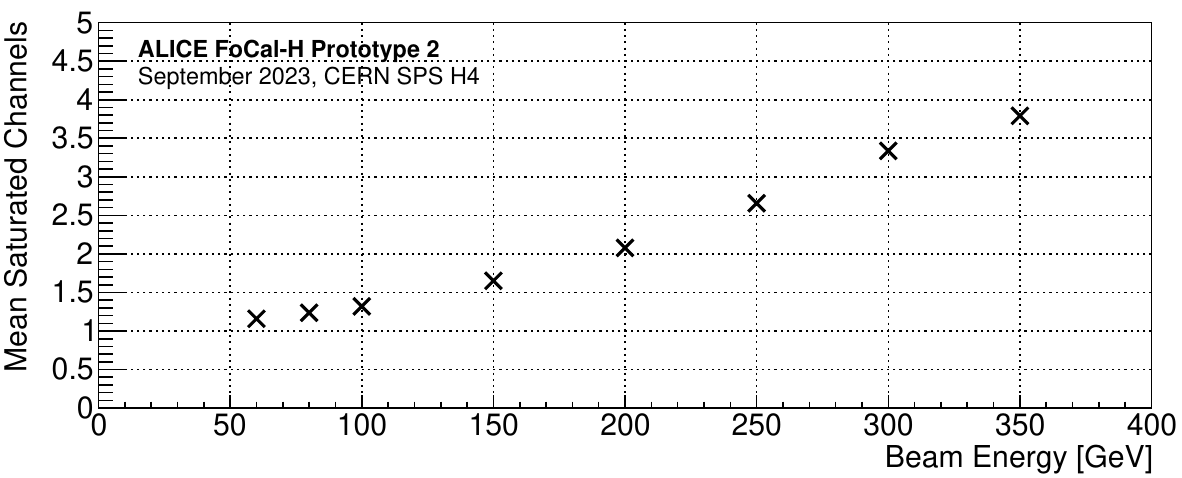}
        \caption{The mean number of saturated HG channels per event as a function of the beam energy. SiPM bias voltage set to 55.4~V.}
        \label{fig:en_sat}
    \end{center}
    \end{figure}
    
\begin{figure}[ht!]
    \begin{center}
        \includegraphics[width = 0.8\textwidth]{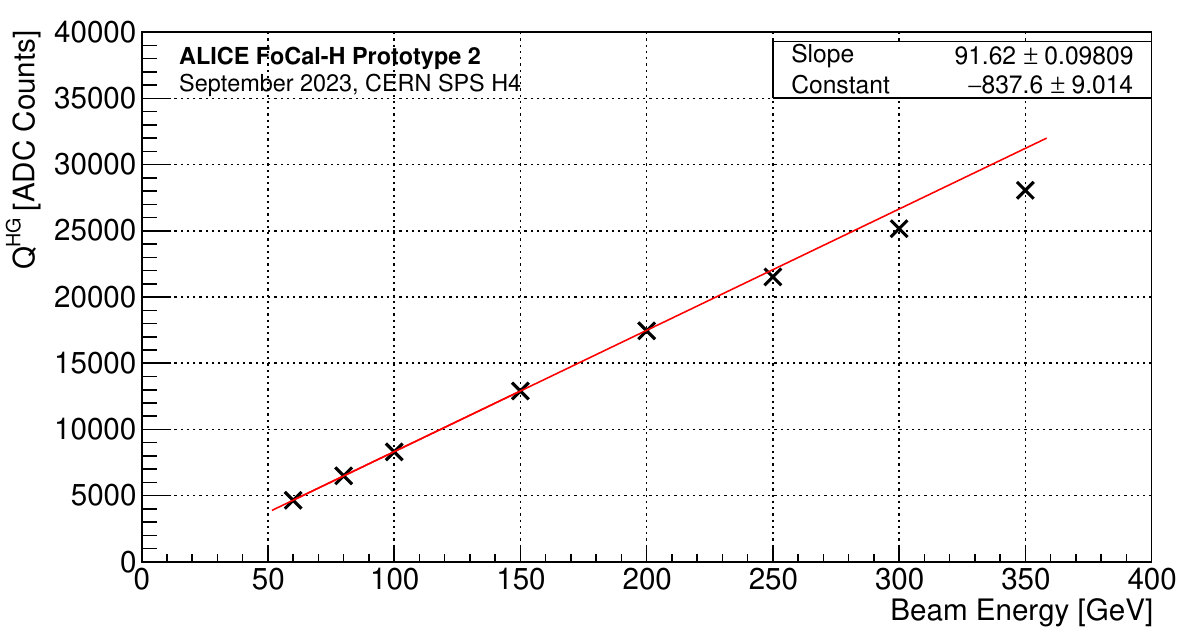}
        \includegraphics[width = 0.8\textwidth]{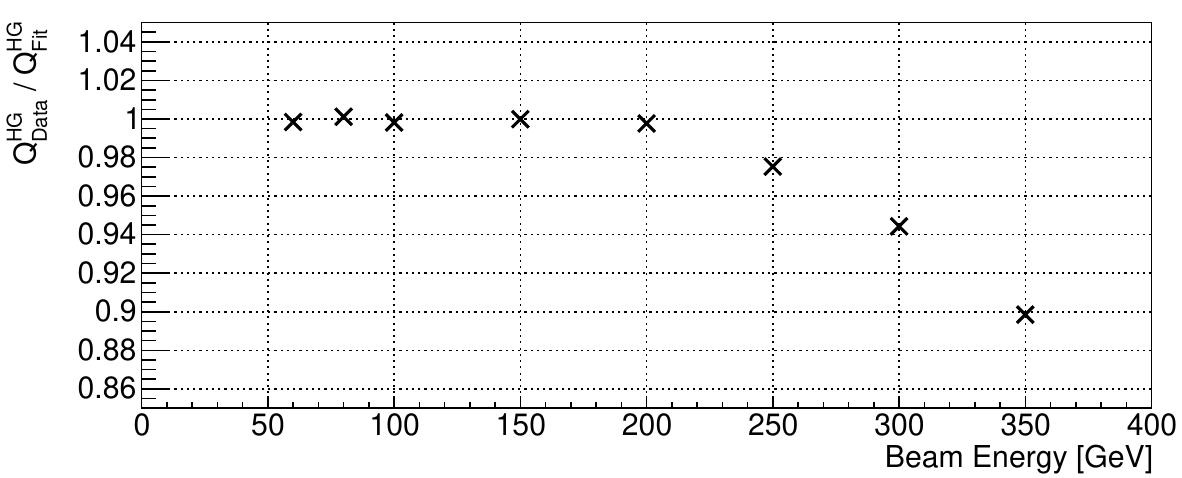}
        \caption{Top panel: Dependence of the mean values of $Q^{\rm HG}$ as a function of the hadron beam energy. Bottom panel: Ratio between the mean values of $Q^{\rm HG}$ and the values predicted by the fit as a function of the hadron beam energy. SiPM bias voltage set to 55.4~V.} 
        \label{fig:lin}
    \end{center}
    \end{figure}

    Due to the limited ADC range some of the channels in an event may 
    saturate (${\rm ADC}_i^{\rm HG}~=~1023$). The mean number of saturated HG 
    channels per event as a function of the beam energy is shown on 
    Figure~\ref{fig:en_sat}. Even for low beam energies
    ($E_{\rm beam}\leq100$~GeV) the mean number of saturated channels is 
    around one and above which indicates that the central channel almost always 
    saturates. The ADC saturation affects the linearity of the charge 
    reconstruction, as shown in Figure~\ref{fig:lin} where the 
    dependence of $Q^{\rm HG}_{\rm event}$ is plotted as a 
    function of the beam energy.
    A linear fit is made using the data points with beam energy from 
    60~GeV to 150~GeV and the line is extrapolated to cover the full energy range.
    The bottom plot shows the ratio between the reconstructed charge and 
    the expected charge (from the linear fit). 
    It can be seen that for beam energies of 250~GeV and higher the departure 
    from linearity is larger than 2\% and even reaches 10\% at 350~GeV.

    The saturation problem is addressed by adopting a charge 
    reconstruction method using a combination of the LG and 
    the HG channels (chapter~\ref{sec:Setup}). 
    In the range where a signal is present in both the HG and the LG 
    components of a channel, a calibrating equation is extracted through a 
    linear fit to the correlation between the recorded values for HG and LG.
    This calibration is performed separately for each channel.
    The linear correlation is assumed to be 
    respected in the range beyond HG saturation and is extrapolated.

\begin{figure}[ht!]
    \begin{center}
        \includegraphics[width = 0.8\textwidth]{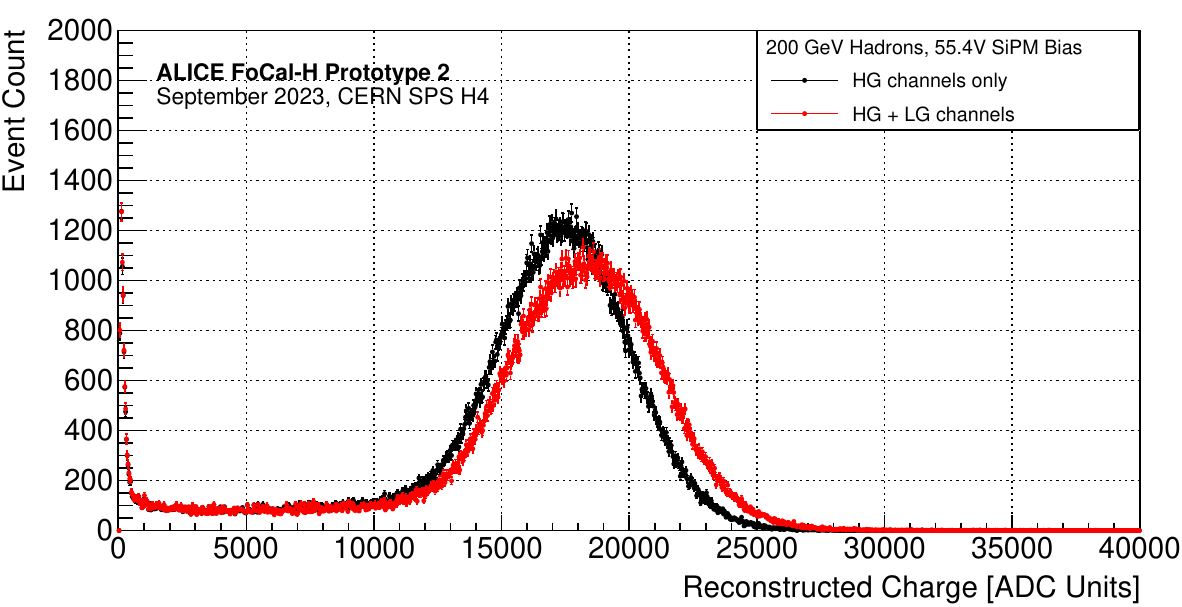}
        \caption{Comparison between reconstructed charge distributions using $Q^{\rm HG}_{\rm event}$
            and $Q^{\rm MIX}_{\rm event}$, showing the resultant shift towards
            larger reconstructed charge when 
            $Q^{\rm MIX}_{\rm event}$ is used. Distributions are normalised
            by the total number of events.
        }
        \label{fig:HG-vs-Cal}
    \end{center}
    \end{figure}

    A new ``combined'' ADC value ${\rm ADC}_i^{\rm MIX}$ is defined as  
        \begin{equation}
            {\rm ADC}_i^{\rm MIX} = {\rm k}_i * {\rm ADC}_i^{\rm LG} + {\rm c}_i
        \end{equation}   
    when an HG channel is saturated and data in the corresponding LG channel is available (with $k_i$ and $c_i$ being the channel-by-channel calibration parameters), and
        \begin{equation}
            {\rm ADC}_i^{\rm MIX} = {\rm ADC}_i^{\rm HG}
        \end{equation}
    when the HG channel is not saturated, or there is no available LG channel data. The final event charge is calculated as
        \begin{equation}
            Q_{\rm event}^{\rm MIX} = \sum^{\rm N}_i{{\rm ADC}_i^{\rm MIX}}.
        \end{equation}
    Thus, whenever LG data is available, and the corresponding HG channel is saturated, we calculate a rescaled ADC value for the HG charge from the LG data, and use that for our reconstruction, effectively increasing the dynamic range. The effect of such a calibration can be seen in Figure~\ref{fig:HG-vs-Cal} where the distribution of ($Q^{\rm HG}_{\rm event}$) is compared to the distribution of
    $Q_{\rm event}^{\rm MIX}$ for 200~GeV. Usage of the rescaled LG values instead of the saturated HG values induces a shift of the total reconstructed charge by about 4.7\%. The relative resolution $\sigma(Q)/Q$ increases from 14.8\% for $Q^{\rm HG}_{\rm event}$ to 16.6\% for
    $Q_{\rm event}^{\rm MIX}$.

    \begin{figure}[ht!]
    \begin{center}
        \includegraphics[width = 0.8\textwidth]{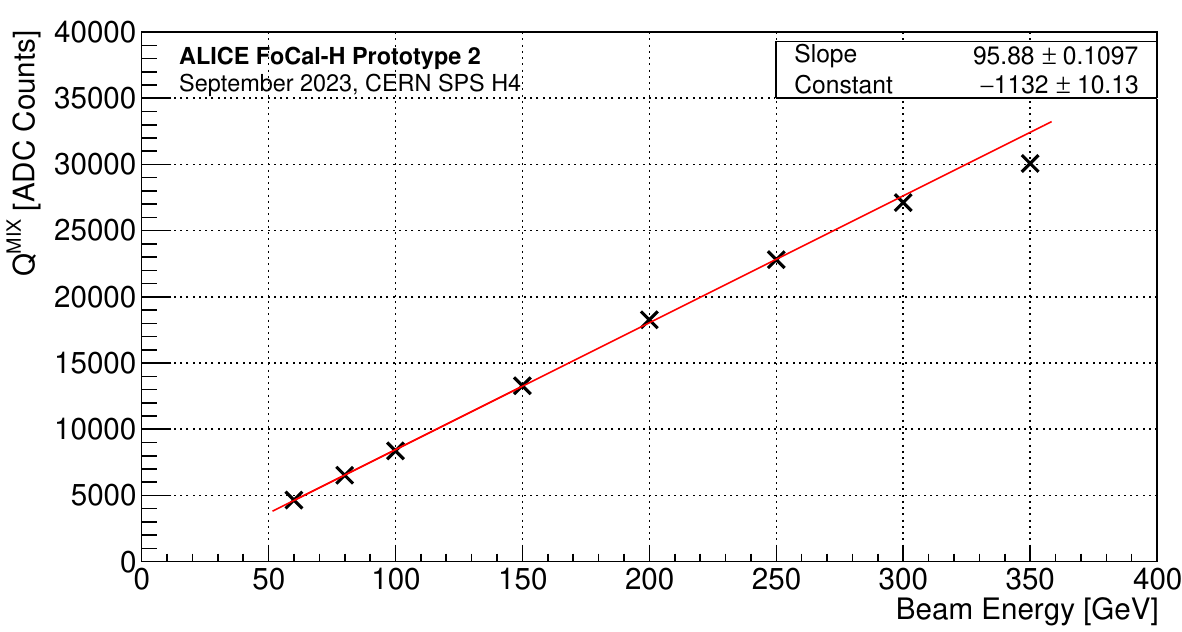}
        {\includegraphics[width = 0.805\textwidth]{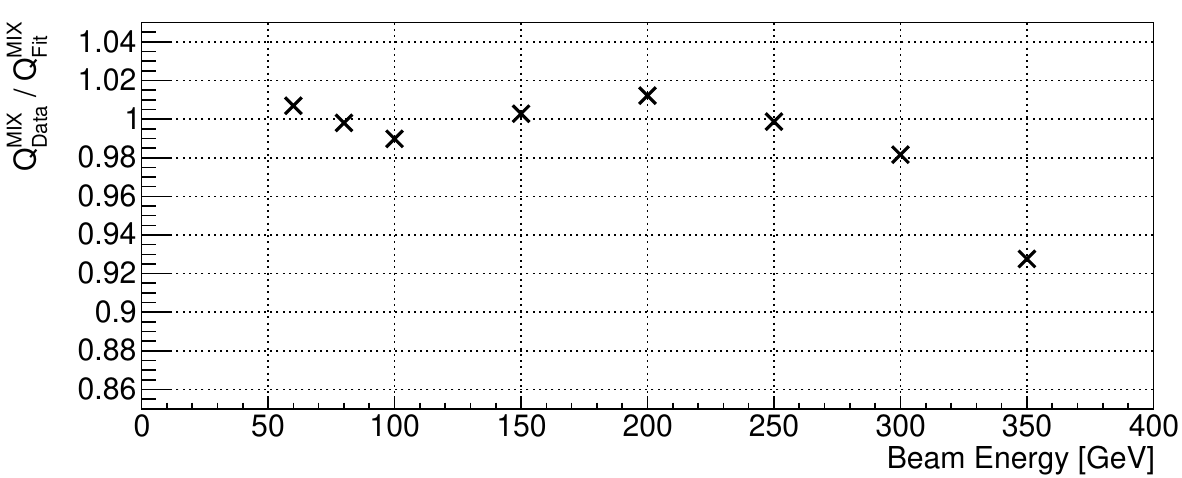}\hspace{-0.5 mm}}
        \caption{Top panel: Dependence of the mean values of $Q^{MIX}$ as a function
            of the hadron beam energy. The linear fit is performed in the interval 60 GeV to 150 GeV and extrapolated to 350 GeV.  Bottom panel: Ratios between the mean values of $Q^{MIX}$ and the values predicted by the fit as a function of the hadron beam energy. SiPM bias voltage set to 55.4~V.}
        \label{fig:lin-cal}
    \end{center}
    \end{figure}

    The dependence of the mean of $Q^{\rm MIX}_{\rm event}$ as a function of the 
    beam energy is presented on Figure~\ref{fig:lin-cal} (top).
    Again, a linear fit is performed for the energy range 60~GeV to 
    150~GeV, and the function is extrapolated to 350~GeV.
    In the ratios between the reconstructed $Q_{\rm event}^{\rm MIX}$ and 
    the fit prediction shown on Figure~\ref{fig:lin-cal} (bottom)
    the departure from linearity reaches 2\% at the 300~GeV data point
    compared to 250~GeV when only using HG. 
    A loss of linearity is still observed even though the LG channels do not saturate. This could be due to transversal and longitudinal 
    shower leakage
    due to the finite dimensions of the FoCal-H prototype and 
    to saturation 
    before the digitization stage at the VMM.

\begin{figure}[ht!]
    \begin{center}
        \hspace{-6.5 mm}
        \includegraphics[width = 0.765\textwidth]{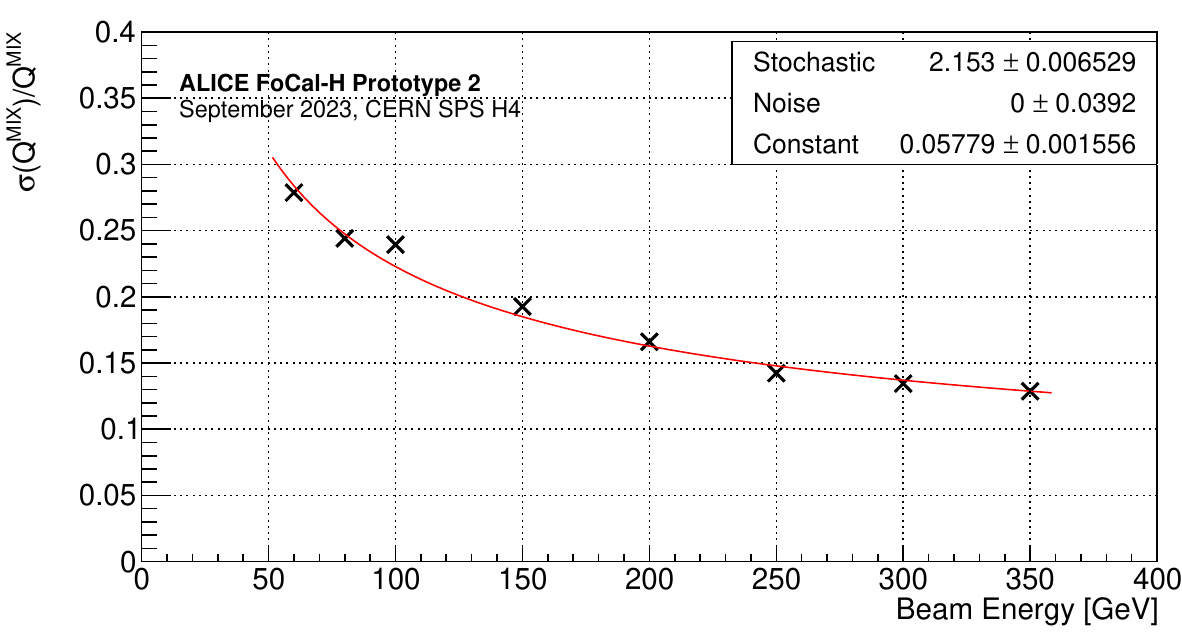}
        \caption{Dependence of the energy resolution obtained with 
        LG/HG calibration on the hadron beam energy.
        SiPM bias voltage set to 55.4~V. The data is fit using Equation~\ref{eq:resolution}.}
        \label{fig:reso}
    \end{center}
    \end{figure}

    While not being the main focus of the presented study, 
    the dependence of the relative energy resolution
    ($\sigma(Q_{\rm event}^{\rm MIX})/Q_{\rm event}^{\rm MIX}$) is presented
    in Figure~\ref{fig:reso} as a function of the beam energy.
    
    The energy dependence of the resolution is approximated using
        \begin{equation}
        \begin{centering}
            \frac{\sigma(Q)}{Q} = \frac{A}{\sqrt{E}} \oplus \frac{B}{E} \oplus C
        \end{centering}
        \label{eq:resolution}
        \end{equation}
    where $A$ is the term describing stochastic effects, $B$ is the noise term
    relating to the readout electronics and $C$ is the constant term 
    relating to shower leakage and miscalibration effects. 
    The obtained values for the resolution parameters are 
        \begin{equation}
            A = (2.15~\pm~0.01) \sqrt{\rm GeV}, ~~~~ B = 0~{\rm GeV}, ~~~~ C = (5.8 \pm 0.2)~\%. 
        \end{equation}
    The stochastic term is larger compared to previously tested readout systems 
    \cite{bib:FoCal_Performance} which can be attributed to the common bias voltage 
    applied to all SiPMs. The noise term is consistent with zero.

\section{Conclusion}
\label{sec:Concusion}
    While primarily 
    intended for the readout of gas detectors, 
    the VMM ASIC can be 
    adapted
    for usage with silicon photomultipliers. 
    The presented results confirm its applicability 
    for calorimetry where a higher dynamic range is required 
    and larger number of channels are active within a single event.
    The VMM dynamic range imposes restrictions on the measurable energy range and resolution, but
    by splitting the signal charge into two VMM channels with
    a charge division of 16:1 the effective dynamic range is
    extended accordingly.
    This principle was first implemented in a SiPM adapter
    board for the VMM  
    with which the initial effects of channel 
    saturation with single gain were
    successfully mitigated. 

    While the developed prototype adapter does not yet allow for 
    individual adjustment of the bias voltage for each detector channel,
    an upgraded adapter 
    could 
    add functionality for
    individual channel tuning and current monitoring.

\section*{Acknowledgements}
    We would like to express our most sincere gratitude to the collaborating individuals, groups and organizations for their invaluable contributions to the success of the test beam campaigns mentioned in this document. We extend our appreciation to the staff and operators of the PS and SPS test beam facilities at CERN for their assistance in setting up and conducting the experiment operations. I.~Bearden, A.~Buhl, L.~Dufke and S. Jia acknowledge support from the The Carlsberg Foundation (CF21-0606) and the Danish Council for Independent Research/Natural Sciences. V.~Kozhuharov and R.~Simeonov acknowledge that partially this study is financed by the European Union-NextGenerationEU through the National Recovery and Resilience Plan of the Republic of Bulgaria, project SUMMIT BG-RRP-2.004-0008-C01. V.~Buchakchiev acknowledges support from ESA through contract number 4000142764/23/NL/MH/rp.

\printbibliography
        
\end{document}